# Tunneling Magnetoresistance and Spin-Dependent Diode Performance in Fully Epitaxial Magnetic Tunnel Junctions with Rock-salt Type ZnO/MgO Bilayer Tunnel Barrier

Hidekazu Saito,[*] Sai Krishna Narayananellore,[†] Norihiro Matsuo,[‡] Naoki Doko,[‡] Shintaro Kon,[‡] Yukiko Yasukawa,[‡] Hiroshi Imamura, and Shinji Yuasa

*National Institute of Advanced Industrial Science and Technology (AIST), Spintronics Research Center, Umezono 1-1-1, Central 2, Tsukuba, Ibaraki 305-8568, Japan*

[‡]*Chiba Institute of Technology, 2-17-1 Tsudanuma, Narashino, Chiba 275-0016, Japan*

**Abstract**

We fabricate fully epitaxial Fe/ZnO/MgO/Fe magnetic tunnel junctions (MTJs) with a bilayer tunnel barrier, in which ZnO has a metastable rock-salt crystal structure. We observe a high magnetoresistance ratio up to 96% at room temperature (RT) and find that these MTJs have asymmetric current-voltage characteristics, and their rectifying performances are largely dependent on the magnetization alignments of the Fe electrodes. Diode responsibilities at a zero-bias voltage ($\beta_0$), which is an important performance index for harvesting applications, are observed up to 1.3 A/W at RT in the antiparallel alignment of the magnetizations while maintaining rather low resistance-area ($RA$) products (a few tens of k$\Omega\mu$m$^2$). Even with the same top and bottom electrodes (Fe), the obtained $\beta_0$ values are comparable to those of reported high-performance tunnel diodes consisting of amorphous bilayer tunnel barriers with polycrystalline dissimilar electrodes. This strongly suggests that the epitaxial ZnO/MgO bilayer tunnel barrier is effective for enhancing the $\beta_0$ without significant increase in the $RA$. In addition, we demonstrate that a zero-bias anomaly in the tunnel conductance, which originates from the magnon excitations at the Fe/barrier interfaces, plays a crucial role in observed spin-dependent diode performance. The results indicate that a fully epitaxial MTJ with a bilayer tunnel barrier is a promising candidate to establish a high-performance high-frequency rectifying system.

PACS: 73.40.Ei, 73.40.Rw, 84.40.-x, 85.75.-d

*Corresponding author.

h-saitoh@aist.go.jp

†Present address: Research Center for Magnetic and Spintronic Materials, National Institute for Materials Science (NIMS), 1-2-1 Sengen, Tsukuba 305-0047, Japan

# I. INTRODUCTION

Metal/insulator/metal (MIM) tunnel diodes have great potential for high-frequency rectifier systems, such as for THz/infrared detectors and radio-frequency energy-harvesting applications [1–17], for which typical semiconductor-based devices cannot operate thanks to their femtosecond electron transition time across the tunnel barrier [18, 19]. For practical applications, the tunnel diode is required to simultaneously have high rectification performance and low junction resistance. The latter is needed to achieve good impedance between the junction and antenna for efficient power transfer, where the typical antenna resistance is on the order of a few hundred ohms [1, 6, 7]. However, there is a trade-off relation between rectification performance and junction resistance, namely, a low-resistance junction tends to have poor rectification performance [7, 8]. This makes it difficult to develop a high-performance tunnel diode that is applicable to practical high-frequency rectifier systems. In most cases, the rectification function in an MIM diode has been achieved using dissimilar electrodes with different work functions. In addition to the dissimilar electrodes, a bilayer tunnel barrier (MIIM structure) has recently attracted much attention, in which the two insulators have different potential barrier heights ($\phi$) [7–9, 20–25]. It has been theoretically shown that the bilayer tunnel barrier is superior to the single tunnel barrier in satisfying both requirements [7–9]. However, it is not clear if the bilayer tunnel barrier is really effective for improving the rectification performance, especially at a zero bias-voltage which is a desirable operating condition from a low-power-consumption point of view. This is because most of the MIIM diodes also have dissimilar electrodes, the same as MIM ones, which makes it experimentally difficult to distinguish between the contributions from the bilayer tunnel barrier and dissimilar electrodes to the rectification performance. To clarify the effect of the bilayer tunnel barrier on rectification, detailed investigations on the rectification performance of an MIIM diode with the same material electrodes are needed. Furthermore, amorphous or polycrystalline tunnel barriers have often been used in both MIM and MIIM tunnel diodes, and a single-crystalline tunnel barrier has never been used in spite of its high potential for tunnel-device applications.

Zinc oxide is a typical oxide semiconductor having a wurtzite (WZ)-type crystal structure under ambient conditions. This material shows a phase transition from WZ- to rock-salt (RS)-type crystal structures under high pressure (~ 9 GPa) [26, 27]. Several groups have successfully grown RS-ZnO(001) thin films on MgO(001) substrate or a templating layer [28–31], where MgO is an insulator with a band gap ($E_g$) of 7.8 eV [32] with an RS-type crystal structure. Since RS-ZnO has a much narrower $E_g$ (~ 2.5 eV [33]) than that of MgO, an RS-ZnO(001)/MgO(001) bilayer can be a promising candidate as a single-crystalline tunnel barrier for an MIIM diode. This is necessary for growing a high-quality RS-ZnO tunnel barrier to conduct a systematic study on its structural properties.

On the other hand, a tunnel junction with ferromagnetic (FM) electrodes is known as a magnetic tunnel junction (MTJ) [34–36]. In an MTJ, the magnetoresistance (MR) ratio is one of the most important performance indexes and defined as ($R_{AP}$ - $R_P$)/$R_P$, where $R_P$ and $R_{AP}$ are the junction resistances between the two FM electrodes with parallel (P) and antiparallel (AP) alignments, respectively. The giant tunneling MR (TMR) effect (> 100%) in MTJs that have single-crystalline or epitaxial tunnel barriers, such as MgO(001) [37–40] and spinels(001) [41–43], makes spintronic practical applications, such as read heads of hard disk drives and magnetoresistive random access memory [39], possible. The giant TMR effect originates from so-called spin-polarized coherent tunneling, in which the coherency of the electron wave function is conserved during tunneling [44–47]. In an epitaxial Fe(001)/MgO(001)/Fe(001) MTJ, highly spin-polarized electrons ($\Delta_1$ Bloch state of the majority-spin band) of Fe(001) dominantly tunnel through MgO(001) as a result of coherent tunneling, which causes the giant TMR effect [37–39, 44–46]. This tunneling effect is not expected to occur in an MTJ with an amorphous tunnel barrier because the electrons with various Bloch states have finite tunneling probabilities (incoherent tunneling) [39]. The giant

TMR effect based on coherent tunneling has also been theoretically predicted in fully epitaxial Fe(001)/RS-ZnO(001)/Fe(001) [28, 48]. Uehara *et al.* fabricated (001)-oriented CoFeB/RS-ZnO/MgO/CoFeB films with an MIIM structure (polycrystalline) and observed sizable MR ratios comparable to those of reference sample with a single MgO barrier [28]. However, their electrical transport properties have not been reported, except the MR ratio near a zero bias-voltage, and reporting on diode performance based on nonlinear and/or asymmetric current-voltage (*I-V*) characteristics has been limited to MTJs with a MgO(001) single tunnel barrier [16, 17].

In this paper, we report on the structural, magneto-transport, and spin-dependent diode properties of Fe/ZnO/MgO/Fe MTJs with a MIIM structure. Note that, for these MTJs, the operation principle of the rectifying process does not come from the spin-torque diode effect, [49, 50] but from the intrinsic nonlinear and/or asymmetric *I-V* characteristics of the MTJ.

## II. EXPERIMENTAL PROCEDURE

MTJ films were grown by molecular beam epitaxy. As illustrated in Fig. 1, each MTJ film consisted of a Au (5 nm) cap / Co (10 nm) pinned layer / Fe (5 nm) top electrode / ZnO (1.2 nm) tunnel barrier / MgO (1.0 nm) tunnel barrier / Fe (30 nm) bottom electrode / MgO buffer (5 nm) layer on a MgO(001) substrate. The source materials are evaporated using electron-beam guns (Fe, ZnO, and MgO) and Knudsen-cells (Au and Co), respectively. For growing the oxide layers, single-crystal ZnO and MgO blocks were used as source materials. The thicknesses of each layer were monitored using a quartz crystal microbalance. Prior to growth, the MgO(001) substrate was heated at 800ºC for surface cleaning. The MgO buffer layers and Fe bottom electrodes were then grown on the substrate at 100ºC, followed by *in situ* annealing at 250°C for 10 min to improve the crystal quality and surface morphology of the Fe bottom electrodes. The MgO tunnel barriers were grown on the Fe bottom electrodes at 100°C. Both the Fe bottom electrodes and MgO tunnel barriers were confirmed to have streaky reflection high-energy electron diffraction (RHEED) patterns, indicating a single-crystalline structure. The ZnO tunnel barriers were then deposited on the single-crystalline MgO(001) tunnel barriers with an $O_2$ pressure of $1 \times 10^{-6}$ Torr. The growth temperature of the ZnO tunnel barriers ($T_{ZnO}$) ranges from room temperature (RT) to 300°C (see Table I). The growth rates of the MgO and ZnO tunnel barriers are 0.01–0.02 nm/s. Finally, the Fe top electrodes and Co pinned and Au cap layers were deposited at RT. The Co layer enhances the coercive force of the Fe top electrode to stabilize the AP state between the magnetizations of the two Fe electrodes. We confirmed through transmission electron microscope (TEM) observations that each layer has excellent uniformity with flat interfaces [51]. For reference, the same MTJ stack without the ZnO tunnel barrier was also fabricated (MgO thickness is 3.0 nm). The films were patterned into tunnel junctions ($3 \times 12$ μm$^2$) using conventional micro-fabrication techniques (e.g., photolithography, Ar ion milling, and $SiO_2$ sputtering). MR curves and diode properties of the MTJs were measured using a conventional DC two-probe method. The magnetic fields were applied parallel to the major axis of the junction corresponding to the easy axis of the magnetization direction of the Fe electrodes.

### III. RESULTS AND DISCUSSION

#### A. Structural analyses

Figures 2 show the RHEED images of the ZnO tunnel barriers (top row of images) and Fe top electrodes (bottom row) of samples A-E. The RHEED images of the ZnO and Fe layers for sample A ($T_{ZnO}$ = RT) exhibited ring patterns (with faint spots for ZnO) [Figs. 2(a) and 2(f)], indicating that both layers are polycrystalline. For samples B, C, and D ($T_{ZnO}$ = 100, 170, and 230°C, respectively), on the other hand, streaky and elongated spotty patterns were respectively observed in the ZnO tunnel barrier [Figs. 2(b), 2(c), and 2(d)] and Fe top electrode images [Figs. 2(g), 2(h), and 2(i)], showing the formation of single-crystalline ZnO and Fe layers. By increasing $T_{ZnO}$ up to 300°C (sample E), broad spots with weak streaky patterns started to appear in the RHEED image of the ZnO layer [Fig. 2(e)]. The positions of the spots do not match the streaks, implying the formation of thermodynamically stable phases such as WZ-ZnO or a ternary compound of a Zn-Mg-O system [29]. For the Fe top electrode, not only streaky patterns but also ring patterns could be observed, meaning that the Fe layer is not single-crystalline. Therefore, samples B, C, and D were confirmed to have a fully epitaxial structure. The observed RHEED patterns of the ZnO tunnel barriers are listed in Table I.

Figure 3(a) shows a cross-sectional high-angle annular dark field scanning TEM (HAADF-STEM) image of sample D. It is clear that the tunnel-barrier part of this MTJ clearly separates into two distinct layers of ZnO and MgO. No intermixing at the interfaces among the tunnel barriers and Fe electrodes could be observed. This result strongly suggests that samples A, B, and C, which had lower $T_{ZnO}$ than that of sample D, also had ZnO/MgO bilayer tunnel barriers without significant intermixing.

Figures 3(b) and (c) show the electron nanobeam diffraction (NBD) patterns of the Fe top electrode and around the ZnO layer, respectively. The diameter of the electron beam spots is ~ 2 nm, so that the NBD patterns in Fig. 3(c) reflect not only the crystal structure of the ZnO layer but also those of the Fe top electrode and MgO tunnel barrier. The NBD patterns of the top Fe electrode revealed a Fe(001)[110] orientation [Fig. 3(b)]. This crystal orientation was confirmed to be identical to that of the Fe bottom electrode. As given in Fig. 3(c), all the spots could be assigned as those of an RS-type crystal structure with a (100)[100] orientation except those corresponding to the Fe top electrode. This indicates that the ZnO tunnel barrier has an RS-type structure with the same crystal orientation as the MgO layer since the observed NBD patterns only consist of the Fe, ZnO and MgO layers. Consequently, the out-of-plane and in-plane crystal orientations among the Fe, ZnO and MgO layers were

7 / 34

determined as top Fe(001) || ZnO(001) || MgO(001) || bottom Fe(001) and top Fe[110] || ZnO[001] || MgO[001] || bottom Fe[110], respectively. Note that the unit cell of the Fe top electrode was turned 45° with regards to the ZnO tunnel barrier, which corresponds to the preferable crystal orientations for the spin-polarized coherent tunneling predicted from *ab initio* calculations [28, 48].

From the HAADF-STEM image, the in-plane lattice mismatch ($\Delta a/a$) between the ZnO tunnel barrier and Fe electrodes was 4.3%, where the lattice constant of Fe was smaller. This observed $\Delta a/a$ is considerably smaller than the expected values (5.3–5.7%) from bulk bcc Fe ($0.286 \times 2^{1/2}$ nm) and RS-ZnO under ambient conditions (0.427–0.428 nm [27, 31, 52]), indicating that the ZnO layer has compressive strain. Using the elastic constants of RS-ZnO [31], the equivalent hydrostatic pressure applied to the ZnO layer was estimated to be 1.4–1.8 GPa, which is close to the critical transition pressure from WZ- to RS-phases (~ 2 GPa [27]), implying that the thickness of the ZnO tunnel barrier (1.2 nm) is near critical thickness while maintaining a RS crystal structure under the experimental growth conditions.

B. Magnetoresistance measurements

Typical MR curves of samples A–E at 20 K and RT with a low bias-$V$ of 10 mV are plotted in Figs. 4(a)–4(e). All the MTJs exhibited a clear TMR effect, i.e. the sizable difference in the junction resistance between the P and AP states. The observed MR ratios are summarized in Table I, together with the resistance-area products in the P state ($R_P A$). The highest MR ratios (127% at 20 K and 96% at RT) were observed for sample D ($T_{ZnO}$ = 230°C) [see also Fig. 4(f)]. To investigate the transport mechanism of the MTJs, relative changes in the MR ratio and junction resistances ($R_P$ and $R_{AP}$) from 20 K to RT are shown in Figs. 4(g) and 4(h), respectively. For epitaxial semiconducting barriers, such as ZnSe [53] and WZ-$Zn_xMg_{1-x}O$ ($x$ = 0.77) [54], the MR ratio and junction resistances tend to rapidly decrease with increasing temperature ($T$). This results in, for example, an absence of the TMR effect at RT. The large reductions in the MR ratio and junction resistances may be caused by the thermally excited carriers or hopping conduction through localized states in the semiconducting barrier at elevated $T$ [53, 54]. For the Fe/ZnO/MgO/Fe MTJs, however, the relative changes in the MR ratios and junction resistances were rather small, up to 42% for the MR ratio, 22% for $R_\text{P}$, and 35% for $R_\text{AP}$. These values are comparable to those for MTJs with conventional insulators [35–38]. These results strongly suggest that direct tunneling, which is favorable for high-frequency applications, is the dominant transport mechanism in the Fe/ZnO/MgO/Fe MTJs even at RT.

The observed MR ratios are basically higher than those observed in MTJs consisting of an amorphous tunnel barrier with conventional 3$d$-ferromagnetic electrodes, such as Fe, Co, and NiFe (typically, 10 to 30% at RT [35, 36, 55]), even if direct tunneling is the dominant transport mechanism. Recall that the coherency of the electron is not conserved for an amorphous tunnel barrier [39]. This indicates that spin-polarized coherent tunneling is effective for the Fe/ZnO/MgO/Fe MTJs, and the $\Delta_1$ state acts as a major tunneling channel in the P state.

### C. Current-voltage characteristic and energy band profile

We found that the epitaxial MTJs have asymmetric $I$-$V$ characteristics with respect to the polarity of the applied bias-$V$, that is rectifying behavior. The solid line in Figure 5, for example, depicts the data on an epitaxial MTJ (sample B) showing the highest asymmetry as the current density ($J$) versus $V$ plot. The current is larger under the positive bias direction in which the electrons tunnel from the bottom electrode to top electrode. Hereafter, the positive and negative bias are defined as a forward and reverse bias direction, respectively. The current asymmetry ($I^+/I^-$) of the fabricated MTJs are listed in Table II, where the $I^+$ and $I^-$ denote the current under forward and reverse biases, respectively.

There are two mechanisms for determining the asymmetric *I-V* characteristics in an MIIM tunnel junction, resonant tunneling and step tunneling (direct tunneling) [8, 20–24, 56, 57]. The former originates from the formation of the resonant state above the bottom of the conduction band of the tunnel barrier with a lower barrier height near the interface between two tunnel barriers, while the latter is due to the asymmetric effective barrier thickness with respect to the polarity of the applied bias-*V* reflecting from its two-step barrier potential. In samples B-D, the larger current is in the forward bias regime, implying that step tunneling is the origin of the asymmetric *I-V* characteristics. To clarify the origin of the asymmetric *J-V* characteristics, we estimated the energy band profile of sample B from simulating the observed *J-V* characteristics. We used the calculation method developed by Grover and Moddel [8], which is applicable to MIIM tunnel junctions. For simplicity, only a tunneling channel of the $\Delta_1$ state in the majority-spin band was considered; therefore, the simulation was conducted in the P state. Other possible transport processes, such as thermionic emission, hopping conduction, and inelastic tunneling, were ignored. The *J-V* characteristics were then determined by the work function of metal electrodes (4.6 eV for Fe(001) [58]), the electron affinity ($\chi$), relative dielectric constant ($\varepsilon_r$), and effective mass ($m^*$) of insulators. To minimize the uncertainty of the estimation, we first deduced the $\chi$ and $m^*$ of the MgO barrier by analyzing the *J-V* characteristics on a reference Fe(001)/MgO(001)/Fe(001) MTJ. The $\varepsilon_r$ of the MgO barrier was used as the reported value measured in MTJs with a single MgO(001) tunnel barrier [59]. Then, the parameters of the RS-ZnO barrier were determined by simulating the *J-V* characteristics of sample B. As plotted with black circles in Fig. 5, the experimental data can be well reproduced through simulation. The determined parameters are summarized in Table III, and the energy band profile is illustrated in the inset of Fig. 5. It is clear that the $\phi$ of the RS-ZnO barrier is much lower than that of the MgO barrier. This results in a different effective barrier thickness with respect to the bias polarities; therefore, the asymmetric *J-V* characteristics. On the other hand, we could not observe a sudden increase in the current under reverse bias-*V*, indicating an absence of the tunneling channel through the resonant state. These results indicate that step tunneling is a major origin of the observed asymmetric *J-V* characteristics. The absence of the resonant state may be due to a rather thin ZnO barrier (1.2 nm thick).

### D. Spin-dependent diode properties

In Fig. 6, the first (d$I$/d$V$) and second (d$^2I$/d$V^2$) derivatives of the current with regards to bias-*V* are shown in the top and second rows of graphs, respectively. The data were obtained from numerical differentiation(s) of the *I-V* characteristics. Polynomial fitting, which has been commonly used to analyze the *I-V* characteristics of a tunnel diode, was not adopted to avoid obtaining erroneous results [60]. The diode responsibility ($\beta$), which is given by 1/2(d$^2I$/d$V^2$)/(d$I$/d$V$) [5], is given in the third row of graphs. The $\beta$ describes how much DC current is generated per unit of RF power of incident high-frequency on a diode. Positive $\beta$ means that a converted DC current flows in the same current direction under the forward bias direction of the diode. The obtained $\beta$-*V* characteristics are valid when the frequency of the incident electromagnetic wave is below a few THz [7, 61].

There are three characteristic features in the d$I$/d$V$ curves (top row of graphs of Figs. 6). The first feature is that the d$I$/d$V$ for the P state is basically higher than that for the AP state due to the TMR effect. For all the Fe/ZnO/MgO/Fe MTJs, the difference in the d$I$/d$V$ between the P and AP states decreased with increasing bias-*V* with both polarities, which is ordinary behavior in an MTJ [34–38]. The second feature is the strong asymmetry with regards to bias-*V* for the fully epitaxial MTJs [Figs. 6(b), 6(c), and 6(d)]. This suggests that the fully epitaxial structure is preferable for rectifying characteristics. The third feature is the broad dip structure for the AP state at *V* ~ 0 V. This results in notable local maximum/minimum and/or shoulder structures in the $d^2I/dV^2$-$V$ curves of the AP state at around $V = \pm 0.1$ V [Figs. 6(g), 6(h), and 6(i)]. Such features, which are collectively called "zero-bias anomalies", have been reported in several MTJ systems [62–67]. The origin of the zero-bias anomaly at elevated $T$ is the collective excitations of local spins (magnon excitations) at the barrier/FM electrode interfaces [62–66]. The noticeable zero-bias anomaly observed in the fully epitaxial MTJs (sample B, C, and D) is probably related to the atomically sharp Fe/ZnO interfaces.

The $\beta$-$V$ characteristics derived from the d$I$/d$V$ and d$^2I$/d$V^2$-$V$ curves are plotted in Figs. 6(l), 6(m), and 6(n). From the $\beta$-$V$ data, we estimated diode performance indexes of the diode responsibility at $V = 0$ ($\beta_0$) and maximum responsibility ($\beta_{max}$), as listed in Table II. The $\beta_0$ is important for energy-harvesting applications because an external power supply is not allowed. On the other hand, the $\beta_{max}$ is important for detector/sensor applications that can be used under the external DC bias. Even with the same top and bottom electrodes (Fe), the obtained $\beta_0$ and $RA$ are comparable to those of high-performance tunnel diodes consisting of amorphous bilayer tunnel barriers with polycrystalline dissimilar electrodes recently reported by Herner *et al*. ($\beta_0 = 2.2$ A/W and $RA \sim 1$ k$\Omega\mu m^2$ [25]). This strongly suggests that the epitaxial ZnO/MgO bilayer tunnel barrier is effective for enhancing the $\beta_0$ without significant increase in the $RA$.

Note that all the Fe/ZnO/MgO/Fe MTJs exhibited higher $\beta_{max}$ for the AP state than that for the P state while $\beta_0$ hardly depended on the magnetization alignments. To clarify the variation in the magnetization alignments on the $\beta$-$V$ characteristics, the relative changes in $\beta$ between the P and AP states [($\beta_{AP} - \beta_P$)/$\beta_P$] in the forward bias regime are illustrated in the bottom row of graphs of Figs. 6. The ($\beta_{AP} - \beta_P$)/$\beta_P$ take a maximum at finite bias-$V$, and the maximum values observed in each MTJ are plotted in Fig. 7. The ($\beta_{AP} - \beta_P$)/$\beta_P$ for the fully epitaxial MTJs reached 70–90% (corresponding to the data on $T_{ZnO}$ = 100, 170, and 230°C), which are basically higher than those of the non-epitaxial MTJs (40–60%). From an application point of view, the results strongly suggest that the AP state might be superior to the P state, at least for detector/sensor applications, as long as the impedance mismatch is not significant.

The appearance of maximum in the ($\beta_{AP}$ - $\beta_P$)/$\beta_P$ versus $V$ plots at a finite bias-$V$ means that the origin of the observed spin-dependent $\beta$-$V$ characteristics is not solely from the difference in the d$I$/d$V$ between the P and AP states because the difference monotonically decreases with increasing bias-$V$. This indicates that the shape of the d$^2I$/d$V^2$-$V$ curve also plays an important role in determining the spin-dependent $\beta$-$V$ characteristics. Thus, we can conclude that the zero-bias anomaly due to the magnon excitations is also responsible for the observed spin-dependent diode performance, especially in fully epitaxial MTJs. The decrease in ($\beta_{AP}$ - $\beta_P$)/$\beta_P$ below several tens of mV (see the bottom row of graphs of Figs. 6) may be due to a low energy cut-off in the magnon spectrum [63, 64].

Finally, we roughly estimated the cut-off frequency ($f_c$) for the Fe/ZnO/MgO/Fe MTJs. For simplicity, an impedance matching between an MTJ and antenna was ignored. The $f_c$ of a tunnel junction is given by $d$ ($2\pi\varepsilon_0\varepsilon_r RA$)$^{-1}$, where $d$ is the thickness of the tunnel barrier, and $\varepsilon_0$ is the dielectric constant. Then, the $R_PA$ values listed in Table I result in a $f_c$ = 0.2 GHz at RT by assuming $\varepsilon_r$ = 10. In this study, it was difficult to further reduce the $RA$ by decreasing the thickness of the ZnO/MgO tunnel barriers because of the relatively large junction area (36 μm$^2$) with a sizable parasitic resistance (~ 10 Ω at RT), prevents reliable electrical transport measurement for a junction with $RA$ lower than ~ 1 kΩμm$^2$. For a Fe/ZnO/MgO/Fe MTJ, however, it is possible to obtain a lower $RA$, down to 0.5 Ωμm$^2$ [28], corresponding to $f_c$ = 2 THz for $\varepsilon_r$ = 10. The next important step for developing the high-frequency rectifier systems is to characterize the diode properties in the lower $RA$ regime.

## IV. CONCLUSION

We fabricated Fe/ZnO/MgO/Fe MTJs with MIIM structures at various $T_{ZnO}$ and conducted systematic investigations on their structural, magneto-transport, and diode properties. Crystallographic studies, i.e., RHEED, HAADF-STEM, and NBD observations, revealed a fully epitaxial Fe(001)/RS-ZnO(001)/MgO(001)/Fe(001) structure for the samples at $T_{ZnO}$ = 100, 170, and 230ºC. High MR ratios up to 96% (127%) were observed at RT (20 K), indicating the existence of spin-polarized coherent tunneling. We found that the fully epitaxial MTJs exhibited notable asymmetric *J-V* characteristics, the mechanism of which was confirmed to be step tunneling, by analyzing the observed *J-V* data. Even with the same top and bottom electrodes (Fe), the obtained $\beta_0$ and *RA* are comparable to those of high-performance MIIM diodes with dissimilar electrodes, strongly suggesting that the epitaxial ZnO/MgO bilayer tunnel barrier can improve diode performance. We demonstrated that the diode properties, especially for the fully epitaxial MTJs, strongly depend on the magnetization alignment of the Fe electrodes. The observed spin-dependent diode performance was due to the TMR effect as well as the magnon excitations at the interfaces between the tunnel barrier and Fe electrodes. The results strongly suggest that a fully epitaxial MTJ with a bilayer tunnel barrier is one of the important building blocks of a THz/infrared rectifying system.

**Acknowledgments**

We thank Dr. Hiroshi Tsukahara (High Energy Accelerator Research Organization) for valuable discussion on the magneto-transport and diode properties. This work was supported by the Grant-in-Aid for Scientific Research on Innovative Area, "Nano Spin Conversion Science" (Grant No. 26103003).

7277 (2014).

**Figure captions**

FIG. 1 Structure of magnetic tunnel junction (MTJ) stack and growth conditions.

FIG. 2 (a)–(e) Reflective high-energy electron diffraction (RHEED) images of ZnO layers (top row of images) and (f)–(j) Fe top electrodes (bottom row) of samples A-D ([100] azimuth of MgO substrate).

FIG. 3 (a) Cross-sectional high-angle annular dark field scanning transmission electron microscope (HAADF-STEM) image, (b) electron nanobeam diffraction (NBD) patterns observed in periphery of Fe top electrode and (c) ZnO layer of sample D with $T_{ZnO}$ = 230°C ([100] azimuth of MgO substrate). Center positions of incident electron beam for NBD observations are indicated with asterisks in HAADF-STEM image.

FIG. 4 (a)–(e) Magnetoresistance (MR) curves of MTJs at 20 K and room temperature (RT) with bias-voltage of 10 mV. (f) MR ratios at 20 K and RT as function of $T_{ZnO}$. (g) Relative changes in MR ratio [MR(RT)/MR(20 K)] and (h) junction resistances [$R$(RT)/$R$(20 K)] as function of $T_{ZnO}$.

FIG. 5 Experimental (solid line) and simulated (black circles) current density ($J$) - voltage ($V$) characteristics of sample B with $T_{ZnO}$ = 100°C in parallel magnetization (P) state at RT. Inset indicates energy band profile of sample B estimated from simulation. Illustration shows relation between electron-tunneling direction and polarity of applied bias-$V$.

FIG. 6 (a)–(e) Bias-$V$ dependence of d$I$/d$V$, (f)–(j) d$^2I$/d$V^2$, and (k)–(o) diode responsibility ($\beta$) in P and AP states. (p)–(t) Relative change in $\beta$ between P and AP states [($\beta_{AP}$ - $\beta_P$)/$\beta_P$] (bottom row of graphs). All data were measured at RT.

FIG. 7 Maximum $(\beta_{AP} - \beta_P)/\beta_P$ values as function of $T_{ZnO}$.

**Table I**. Sample code, growth temperatures of ZnO ($T_{ZnO}$), RHEED patterns of ZnO tunnel barriers, magnetoresistance (MR) ratios at 10 mV, and resistance-area products in the P state ($R_PA$) at 10 mV for MTJs, respectively.

| Sample | $T_{ZnO}$ (°C) | RHEED of ZnO | MR (%) | | $R_pA$ (kΩμm²) | |
|---|---|---|---|---|---|---|
| | | | 20 K | RT | 20 K | RT |
| A | RT | faint spots + rings | 56 | 40 | 7.5 | 6.4 |
| B | 100 | streaks | 70 | 51 | 18 | 16 |
| C | 170 | streaks | 94 | 54 | 13 | 11 |
| D | 230 | streaks | 127 | 96 | 9.6 | 8.6 |
| E | 300 | streaks + spots | 81 | 51 | 15 | 12 |

**Table II**. Current asymmetry ($I^+/I^-$) at $V = 0.1$ and 0.5 V, diode responsibility at $V = 0$ ($\beta_0$) and maximum responsibility ($\beta_{max}$) in P and AP states for MTJs at RT, respectively.

| Sample | $I^+/I^-$ | | $\beta_0$ (A/W) | | $\beta_{max}$ (A/W) | |
|---|---|---|---|---|---|---|
| | 0.1 V | 0.5 V | P state | AP state | P state | AP state |
| A | 1.1 | 1.2 | 0.4 | 0.3 | 1.5 | 1.9 |
| B | 1.3 | 2.6 | 1.1 | 1.3 | 1.7 | 2.4 |
| C | 1.2 | 2.3 | 0.9 | 1.1 | 1.2 | 2.3 |
| D | 1.1 | 1.7 | 0.5 | 0.4 | 1.1 | 1.5 |
| E | 1.2 | 1.5 | 0.8 | 1.1 | 2.3 | 3.1 |

**Table III.** Electron affinity ($\chi$), relative dielectric constant ($\varepsilon_r$), and effective mass ($m^*$) of MgO and RS-ZnO tunnel barriers determined by simulating *J-V* characteristics observed for sample B. Barrier heights ($\phi$) correspond to energy difference between $\chi$ of tunnel barriers and work function of Fe(001) of 4.6 eV [58]. Nominal thicknesses of MgO (1.0 nm) and ZnO (1.2 nm) tunnel barriers were used for simulation.

| barrier | $\chi$ (eV) | $\varepsilon_r$ | $m^*$ | $\phi$ (eV) |
|---|---|---|---|---|
| MgO | 0.65 | 8.8[a] | 0.10 | 4.0 |
| RS-ZnO | 3.7 | 12 | 0.55 | 0.89 |

[a]Reference [59]

Figure 1

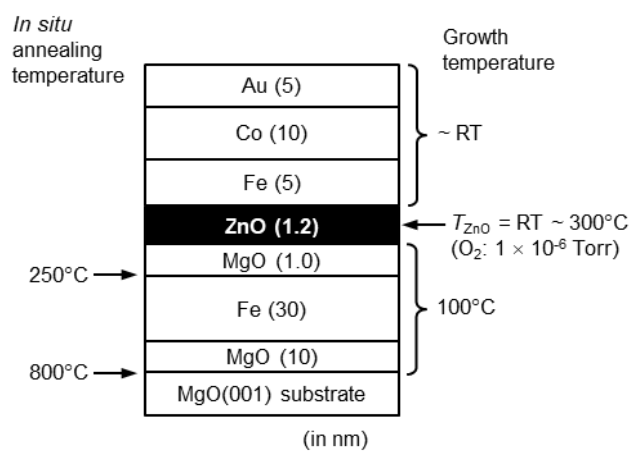

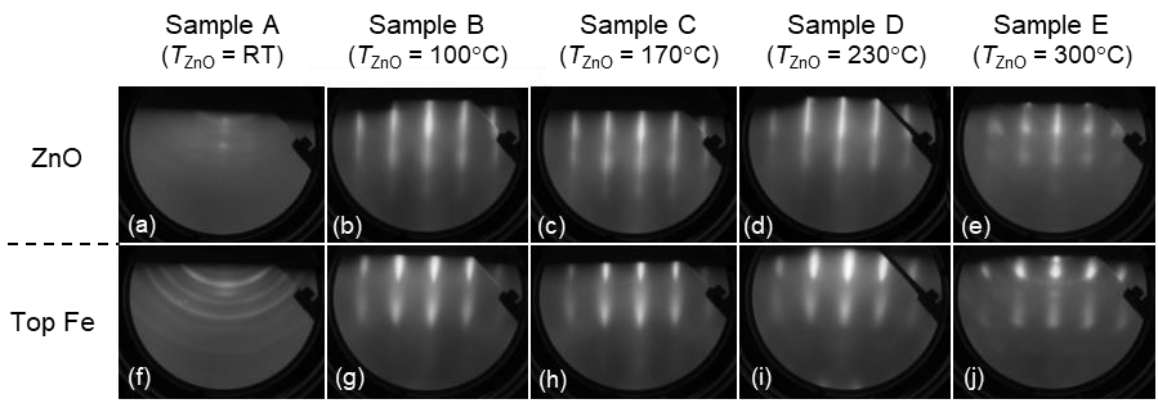

Figures 2(a)-2(j)

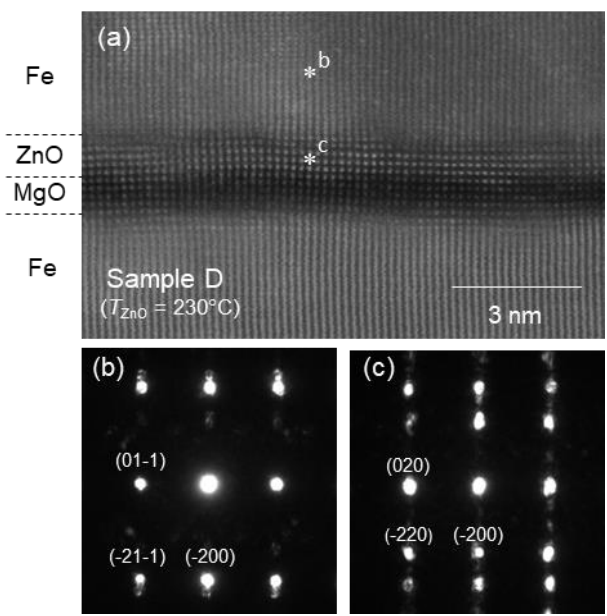

Figures 3(a)-3(c)

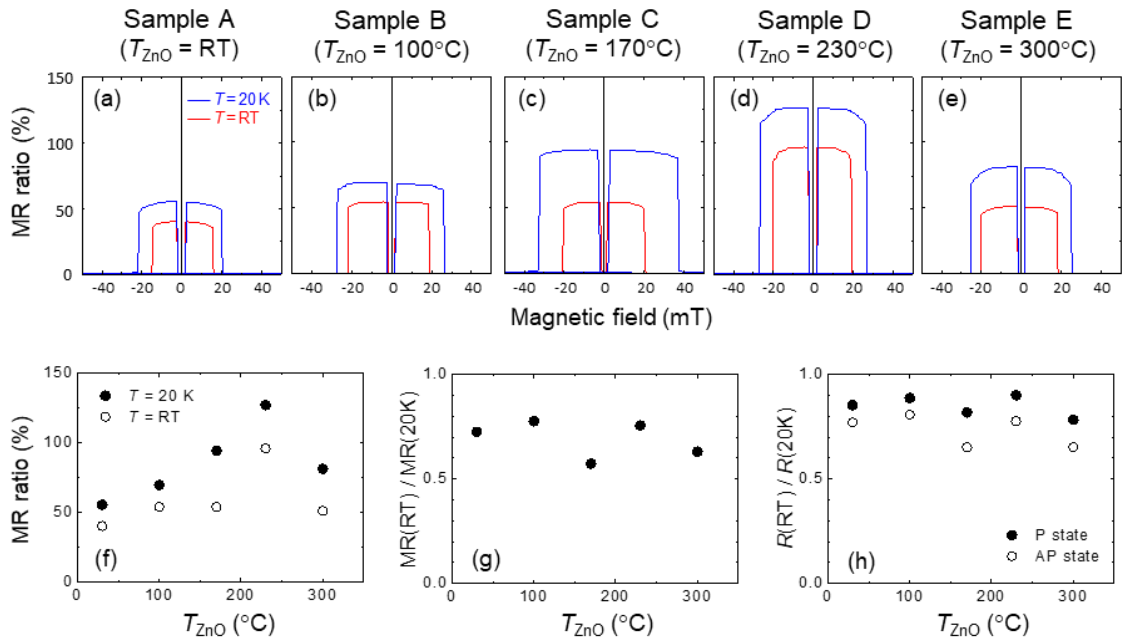

Figures 4(a)-4(h)

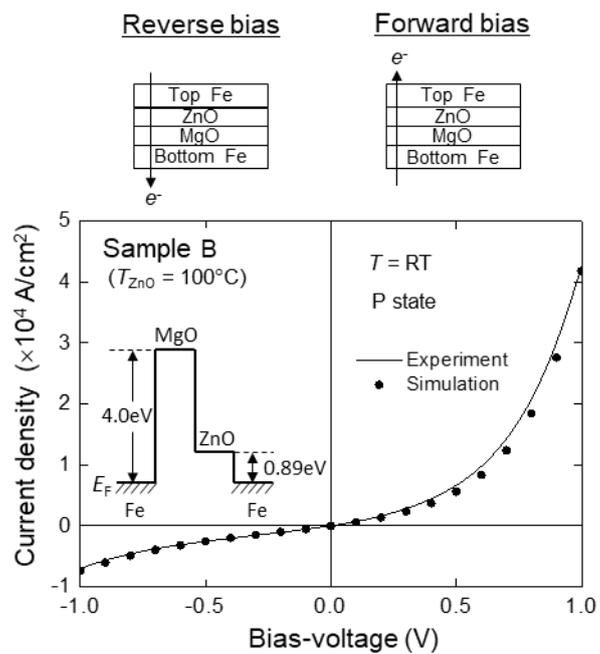

Figure 5

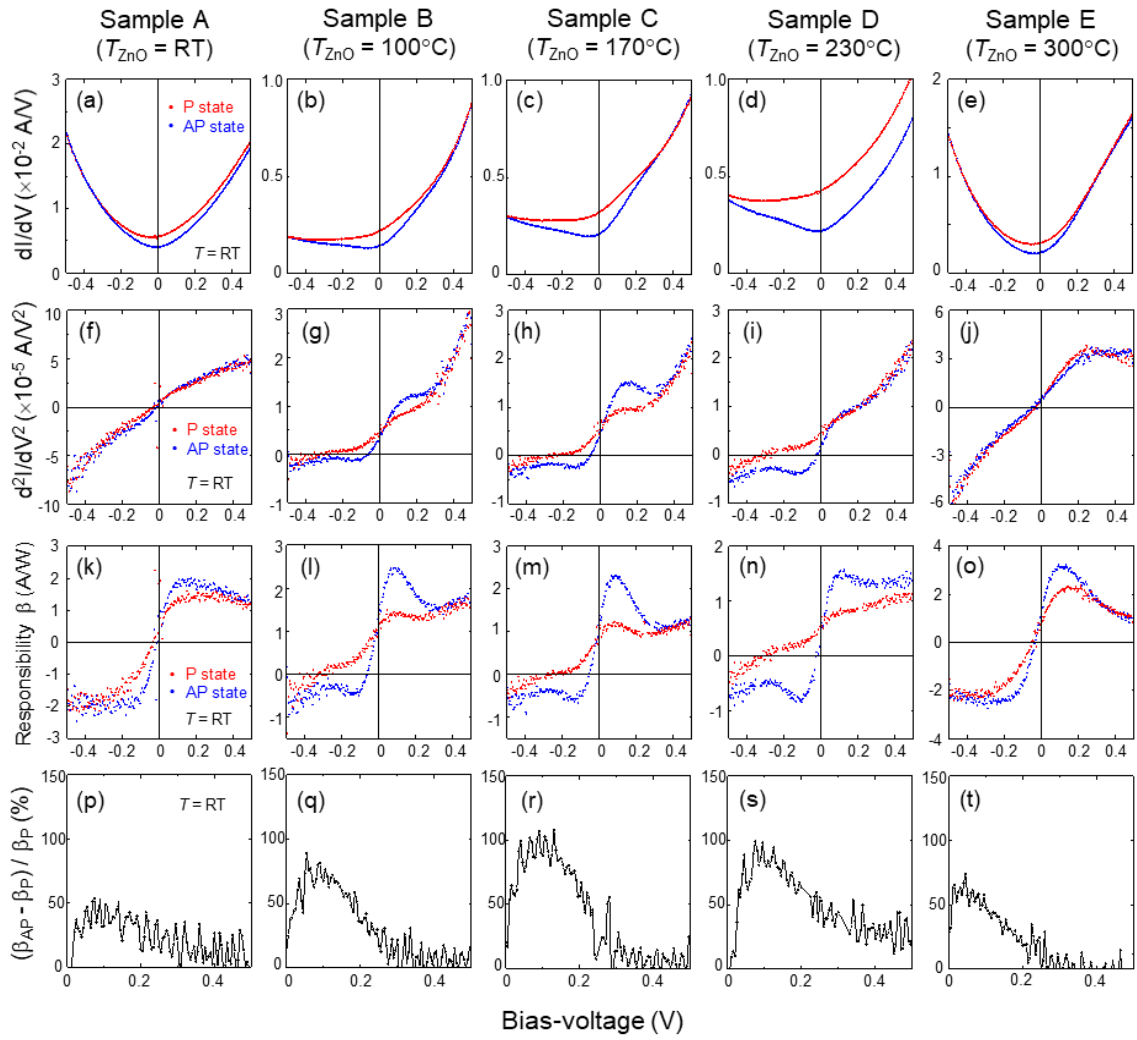

Figures 6(a)-6(t)

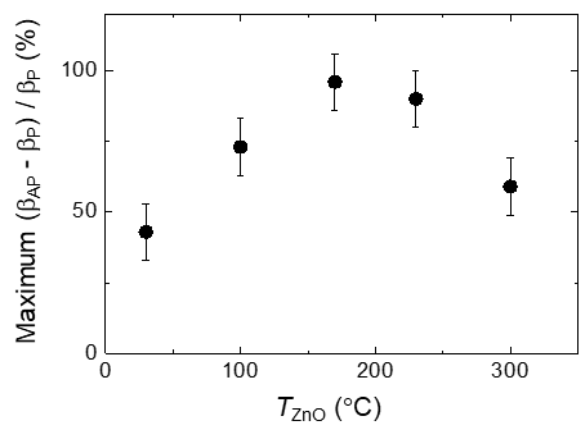

Figure 7